\documentclass[]{spie}  %>>> use for US letter paper
\pdfoutput=1
\usepackage{pbox}
\usepackage{graphicx}
\usepackage{caption}
\usepackage{subcaption}
\usepackage{color}
\usepackage{float}
\usepackage[ruled,vlined]{algorithm2e}
\SetAlFnt{\small}
\usepackage{hyperref}
\usepackage{amsmath}

\DeclareMathOperator*{\argmin}{\arg\!\min}
\newcommand{\amx}{\underset{\boldsymbol{\theta_D}\in \Theta_D }{\operatorname{argmax}}} 
\newcommand{\amn}{\underset{\boldsymbol{\theta_G}\in \Theta_G }{\operatorname{argmin}}}

\newcommand{\degree}{\ensuremath{^\circ}}
\usepackage{txfonts}
\usepackage{amsmath,amsfonts,amssymb}
\usepackage{graphicx}

\title{Neural network based image reconstruction with astrophysical priors}

\author[a]{R. Claes}
\author[a]{J. Kluska}
\author[a]{H. Van Winckel}
\author[b]{M. Min}
\affil[a]{Instituut voor Sterrenkunde (IvS), KU Leuven, Celestijnenlaan 200D, 3001, Leuven (Belgium)}
\affil[b]{SRON Netherlands Institute for Space Research (Netherlands)}

\authorinfo{Further author information: (Send correspondence to Jacques Kluska)\\Jacques Kluska: E-mail: jacques.kluska@kuleuven.be, Telephone: +32 16 32 84 51 }

\begin{document} 
\maketitle

\begin{abstract}
With the advent of interferometric instruments with 4 telescopes at the VLTI and 6 telescopes at CHARA, the scientific possibility arose to routinely obtain milli-arcsecond scale images of the observed targets. Such an image reconstruction process is typically performed in a Bayesian framework where the function to minimize is made of two terms: the data likelihood and the Bayesian prior. This prior should be based on our prior knowledge of the observed source. Up to now, this prior was chosen from a set of generic and arbitrary functions, such as total variation for example. Here, we present an image reconstruction framework using generative adversarial networks where the Bayesian prior is defined using state-of-the-art radiative transfer models of the targeted objects. We validate this new image reconstruction algorithm on synthetic data with added noise. The generated images display a drastic reduction of artefacts and allow a more straightforward astrophysical interpretation. The results can be seen as a first illustration of how neural networks can provide significant improvements to the image reconstruction of a variety of astrophysical sources.
\end{abstract}

% Include a list of keywords after the abstract 

\section{INTRODUCTION}
\label{sec:intro}  % \label{} allows reference to this section
Obtaining an angular resolution in the milli-arcsecond range is vital to study the morphology of a variety of astrophysical targets such as evolved and young stars, binaries, active galactic nuclei or solar system bodies. 
Current and planned single telescope facilities do not reach such an angular resolution, but thanks to the development of optical (visible and infrared) interferometry, these resolutions can now be routinely obtained.

Using this technique one does not directly observe images of the target but instead interferometric fringes that contain information about the Fourier transform of the image at given spatial frequencies $u$ and $v$ which depends on the projected telescope baselines on the sky. 
%An observation at a given baseline measures fringes by its coordinates $u$ and $v$.
By observing with many different baselines,  one can covert the measurements in the well-covered ($u, v$)-plane, to try to reconstruct the image of the target on the sky. % by filling the missing spatial frequencies (or cover the ($u$, $v$)-plane) and performing an inverse Fourier transform.
%Therefore, such process of image reconstruction requires a rich (u, v)-plane that can be achieved by instruments that recombine enough telescopes.
With the advent of new generation of interferometric instruments which can recombine four or more telescopes, such as GRAVITY \cite{GRAVITY}, the Precision Integrated-Optics Near-infrared Imaging ExpeRiment \cite{PIONIER} and the Multi AperTure mid-Infrared SpectroScopic Experiment \cite{MATISSE} at the Very Large Telescope Interferometer (VLTI) or the Michigan Infrared Beam Combiner-X \cite{MIRCX} at the Center for High Angular Resolution Astrophysics (CHARA), we have entered the imaging era in optical interferometry \cite{Kloppenborg2010,Roettennbacher2016,Paladini2018,Kluska2020}.

Data from these instruments have revealed that the observed targets display often complex morphologies that are difficult to retrieve using geometric model fitting, as many parameters ($>10$) are needed to reproduce the data. In this process,  there is a high risk of including a model bias \cite{Lazareff2017,Kluska2019,Perraut2019}. 
The imaging technique is, therefore, unique to unveil complex and unexpected morphology at milliarcsecond scale, provided the reconstruction does not include artifacts which hamper the astrophysical interpretation.
Image reconstruction algorithms are therefore crucial to recover the best quality images.
%We note that similar need of image reconstruction method is needed in radio interferometry with the  

Reconstructing an image from optical interferometric data is typically performed in a Bayesian framework where the image ($\boldsymbol{x}$)  which maximizes the \textit{a posteriori} probability given the observations ($\boldsymbol{y}$) is sought \cite{Thiebaut2017}.

This probability ($P(\boldsymbol{x}\mid\boldsymbol{y})$) can be expressed, using the Bayes rule, by:
\begin{equation}
    P(\boldsymbol{x}\mid\boldsymbol{y}) = \frac{P(\boldsymbol{y}\mid\boldsymbol{x}) P(\boldsymbol{x})}{P(\boldsymbol{y})}, \label{eq:Bayes}
\end{equation}
where $P(\boldsymbol{y}\mid\boldsymbol{x})$ is the likelihood of the data, $P(\boldsymbol{x})$ is the \textit{a priori} distribution of $\boldsymbol{x}$ and $P(\boldsymbol{y})$ is called the evidence.
The evidence is not taken into account in the image optimization procedure, as it does not depend on the image $\boldsymbol{x}$.

In practice, instead of maximising the probability one searches the \textit{maximum a posteriori} solution (i.e., the image with the \textit{maximum a posteriori} probability; $\boldsymbol{x}_{MAP}$) that minimizes the negative logarithm of Eq.\,\ref{eq:Bayes} that is called the cost function ($f$):
\begin{eqnarray}
    \boldsymbol{x}_{MAP} & = & \argmin_{\boldsymbol{x}} f  = \argmin_{\boldsymbol{x}}\big( -\log(P(\boldsymbol{y}\mid\boldsymbol{x})) - \log(P(\boldsymbol{x}))\big) \\
    & = & \argmin_{\boldsymbol{x}} f_\mathrm{data} + \mu f_\mathrm{rgl},\label{eq:Bayes2}
\end{eqnarray}
where $f_\mathrm{data}$ is the data likelihood cost function (e.g. $\chi^2$), $f_\mathrm{rgl}$ is the regularization and $\mu$ is the regularization weight that sets the strength of the regularization.

The regularization influences the image reconstruction by promoting images with a certain brightness distribution, independent of the data likelihood. 
By doing so it determines how the Fourier space is extrapolated between ($u$, $v$)-plane measurements.

Most commonly generic functions coming from the signal processing community are used such as maximum entropy, quadratic smoothness, total variation or Laplacian regularizations \cite{Thiebaut2008,Baron2010,Renard2011,Hofmann2016,Thiebaut2017}.
Based on this Bayesian framework, the regularization should incorporate our prior expectations or astrophysical knowledge of the brightness distribution. This is not the case for these common generic regularizations, as such distributions are too complex to formalize in a simple equation.

%These regularizations are not reflecting the prior astrophysical knowledge or expectation of the observed source intensity distribution as implied by the Bayes equation.  Such expected distribution is complex to formalize in an equation and use as a regularisation. 

We therefore present here a novel image reconstruction framework based on convolutional neural networks (CNNs) \cite{LeCun2015}. 
We employed neural networks trained in a deep convolutional generative adversarial network (GAN) \cite{radford2015unsupervised} to reconstruct images. The method is called ORGANIC: Object Reconstruction using Generative Adversarial Networks for InterferometriC data.
Among other properties, it allowed us to use CNNs as a regularisation, making it learn the prior distributions from images generated by models of astrophysical sources.
The method can be applied to a variety of astrophysical sources if models are available. In this paper, we focus on circumstellar disks.

In section \ref{framework} the image reconstruction framework and neural network architecture is presented. The results obtained on artificial datasets are presented in section \ref{validation}. The conclusions and future prospects are discussed in section \ref{conclusion}.
%equations for the bayesian framework
%\begin{equation}
%    f_{data}(\boldsymbol{x}) \propto -\log(P(\boldsymbol{y}\mid\boldsymbol{x})) 
%    \label{ch2:likelyhood}
%\end{equation}
%\begin{equation}
%    \mu f_{prior}(\boldsymbol{x}) \propto -\log(P(\boldsymbol{x}))
%    \label{ch2:regularisation}
%\end{equation}
%\begin{equation}
%    \min_{x\in\Omega}\{f(\boldsymbol{x})=f_{data}(\boldsymbol{x})+\mu f_{prior}(\boldsymbol{x})\}
%    \label{ch2:minProblem}
%\end{equation}

\section{The ORGANIC image reconstruction framework}\label{framework}
\subsection{Generative adversarial networks}

\begin{figure*}[t]
    \centering
    \includegraphics[width = 17cm]{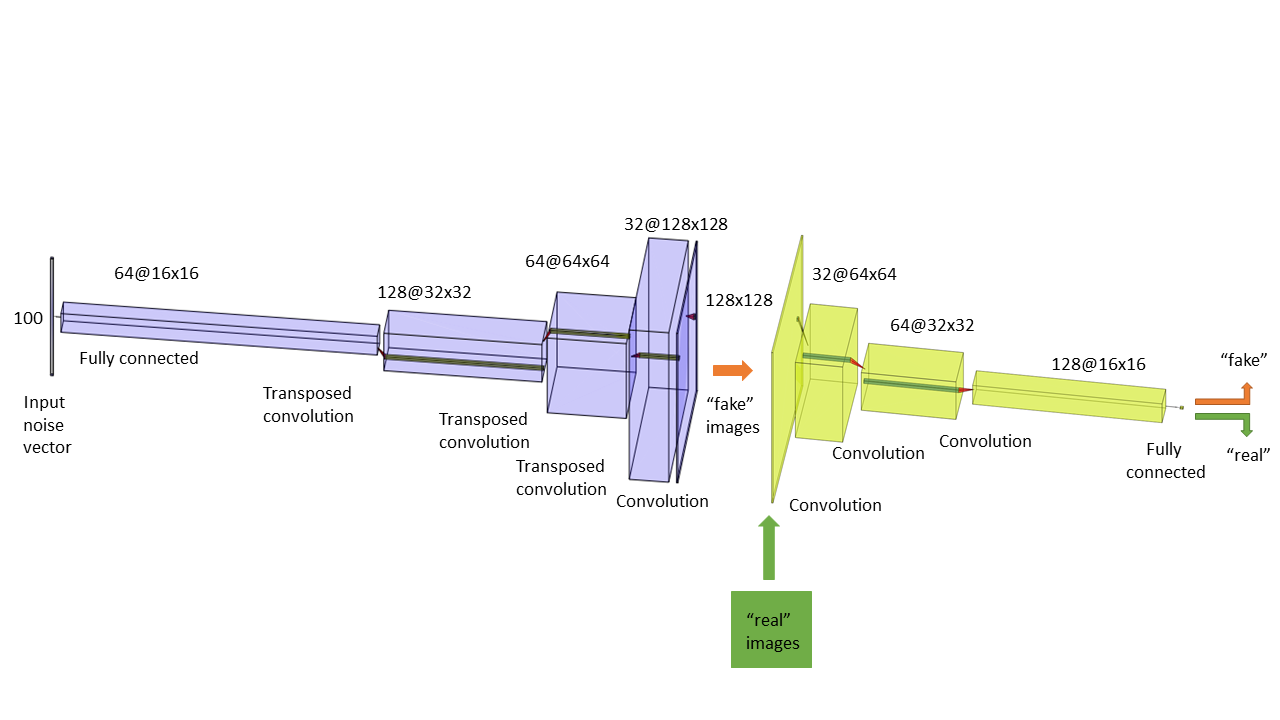}
    \caption{Schematic representation of the used GAN }%\RC{I think the networks are too big to properly illustrate them, although this one looks better IMO, the kernels of the transposed convolution are not corectly displayed, this does however not matter if we do not call them kernels}}
    \label{schemeGan}
\end{figure*}
%\begin{figure*}
 %   \centering
 %   \includegraphics[width = 18cm]{Imagepaper2.png}
 %   \caption{Schematic representation of a GAN \RC{I think the networks are too big to properly illustrate them}}
 %   \label{schemeGan}
%\end{figure*}

A GAN is a combination of two CNNs called the generator and the discriminator (see figure \ref{schemeGan}) which compete against each other \cite{goodfellow2014} .
During the training process the generator learns to generate images that look like the model images, while the discriminator is supplied by both model images and generator images and learns to discriminate between the two by generating a number between 0 (``fake'' image) and 1 (``true'' image).
In other words the discriminator learns to classify images produced by the generator as ``fake'' and those from the training dataset as ``real''. 
Meanwhile, the generator tries to make the discriminator labeling the generator image as ``real''.
Therefore, during the training of a GAN, the discriminator network $D$ and generator network $G$, try to achieve goals in competition of each other with the object function being defined by:
\begin{equation}
    \amn \;\amx V(D,G),
    \label{ganObj}
\end{equation}
with:

\begin{equation}
    V(D,G) = \mathbb{E}_{\boldsymbol{x}\sim P_\mathrm{model}}[\log(D(\boldsymbol{x};\boldsymbol{\theta_D}))] + \mathbb{E}_{\boldsymbol{z}\sim P_{\boldsymbol{z}}}[\log(1-D(G(\boldsymbol{z};\boldsymbol{\theta_G});\boldsymbol{\theta_D}))]
    \label{minmaxProblem}
\end{equation}
and $\boldsymbol{\theta_D}$ and $\boldsymbol{\theta_G}$ represent the trainable parameters describing the discriminator $D$ and generator $G$ respectively. 
$\boldsymbol{x}$ are sample images drawn from a distribution $P_\mathrm{model}$ and $\boldsymbol{z}$ is a noise vector sampled from distribution $P_{\boldsymbol{z}}$.
The global optimal value of $V(D,G)$ is achieved when the distribution of the data created by the generator $P_G(\boldsymbol{x})$ is equal to $P_\mathrm{model}(\boldsymbol{x})$ \cite{goodfellow2014}.

%\begin{equation}
%    V(D,G) = \mathbb{E}_{\boldsymbol{x}\sim P_{data}}[\log(D(\boldsymbol{x};\boldsymbol{\theta_D}))] + \mathbb{E}_{\boldsymbol{z}\sim P_{\boldsymbol{z}}}[\log(D(G(\boldsymbol{z};\boldsymbol{\theta_G});\boldsymbol{\theta_D}))]
%    \label{minmaxProblem2}
%\end{equation},

% This alternative cost function can also be seen to be the commonly used binary cross entropy loss function if the output targets are set to 1. 

%A sketch of the implemented framework is given in figure \ref{schemeGan}. 
%In practice, a noise vector $\boldsymbol{z}$ is provided to the generator, which is trained to convert this input into an image which the generator classifies as real by producing an output close to 1. 
%By doing so it tries to convince the discriminator that the images it produces are part of the real image dataset. 
%The discriminator is supplied with images from both a training dataset representing the distribution $P_ {data}(\boldsymbol{x})$ and the output of the generator, consisting of fake images. For these outputs the GAN is trained to produce outputs close to 1 or 0 respectively. 

\subsection{Training the GAN}
Building a GAN architecture that solves Eq.\,\ref{ganObj} requires to define both $G$ and $D$, an optimization strategy and an adequate training dataset. 
%Solving this problem is prone to non-convergence due to the numerical approach that needs to be used. 
There is, however, no clear consensus on how to best define the network architectures and training routine that will converge. 
Therefore, much of the information presented here has been arbitrarily chosen from common practices in the field of deep learning when possible.

\subsubsection{GAN Architecture}
Both of our neural networks are CNNs, as  they consist of both fully connected and convolutional layers.
CNN have beneficial properties when dealing with image data \cite{Goodfellow-et-al-2016} as they are made with peculiar layers called the convolution layers.
They allow the extraction of meaningful information from the image while limiting the amount of parameters to train. 
The architectures of both the discriminator and generator networks are listed in tables \ref{discrArchitecture} and \ref{genArchitecture} respectively. More information about these layers can be found in appendix \ref{app:layers}

%We describe them in the following paragraphs.
%Two additional concepts are typically applied, combined with this expression, in convolutional layers. 
%These are stride and zero-padding. Stride can be seen as the step size with which the kernel moves over the input images when preforming the convolution. By choosing a stride larger than 1 the output size of output image is smaller than that of the input image. %include the image? No.

%Zero padding refers to the extending of the input images by pixels with a value equal to 0 at the edges of the input images. The most commonly used type of padding is same-padding. In same padding the amount of zeros added to the sides of the images is chosen such that the size of the output images is equal to that of the input images if the used strides are 1, half the size if they are 2 and so on. If no such padding is added the kernel can not be extended over the image boundary and the size output image decreases as a consequence of this. 

\paragraph{The discriminator $D$}

\begin{table*}[t]
\caption{The architecture of the Discriminator network. }
\label{discrArchitecture}
\begin{tabular}{ l | l | p{1.75cm} | p{1.2cm}  |l | l | p{1.5cm} } 
\hline
 layer & activation & Number of \newline kernels &kernel \newline size & stride & output shape & trainable \newline parameters\\
\hline
\hline
2D  convolution & Leaky relu $\alpha = 0.25$ & 32 & $4\times 4$ & (2,2) &  (64,64, 32) & 320\\ 
2D  convolution & Leaky relu $\alpha = 0.25$ & 64 & $4\times 4$ & (2,2) &   (32,32, 64) & 18496\\ 
2D  convolution & Leaky relu $\alpha = 0.25$ & 128 & $4\times 4$ & (2,2) &  (16,16, 128)  & 73856\\ 
Fully connected  with bias & Sigmoid & / & / & / &  1  &32769\\ 

\hline
\end{tabular}
\end{table*}

The discriminator takes an image 128$\times$128 pixels as an input and gives a single value between 0 and 1 as an output.
In the discriminator the leaky ReLu activation function \cite{Relu} with a leak strength of $\alpha = 0.25$ was used for all the convolutional layers. 
In the final, fully connected layer a sigmoid activation function is used, as this restricts the output range to be between 0 to 1. 
This is done as $D(\boldsymbol{x})$ represents the probability that $\boldsymbol{x}$ came from the data rather than $P_G(\boldsymbol{x})$.  
To avoid over-confidence, during both the training of the GAN and the image reconstruction, we used a dropout of 0.3, i.e., each of the convolutional kernels has a 30\% chance of not contributing towards the output of the discriminator and the back propagation of the errors in the discriminator and generator networks. The outputs of the kernels which are not dropped are scaled up by $1/(1 - 0.3)$ such that the sum over all outputs is unchanged.
The discriminator is made of 125441 trainable parameters (see Table\,\ref{discrArchitecture}).

\paragraph{the Generator $G$}

\begin{table*}[t]
\caption{The architecture of the Generator network. }
\label{genArchitecture}
\begin{tabular}{ l | l | p{1.6cm} | p{1.2cm}  |l |l | p{1.6cm}  } 
\hline
 layer & activation & Number of\newline kernels &kernel\newline size & stride & output shape & trainable\newline parameters\\
\hline
\hline
Fully connected without bias & Leaky relu $\alpha = 0.1$  & / &/ & / & 65536 &  6553600\\ 
reshape & / & / &/ & / & (16 ,16 , 256) &\\ 
2D transpose convolution& Leaky relu $\alpha = 0.1$ & 128 & $4\times 4$ & (2,2) &  (32 ,32 , 128)  &   524288\\
Batch normalization & / & / & / & / &  (32 ,32 , 128) & 256\\
2D transpose convolution & Leaky relu $\alpha = 0.1$ & 64 & $4\times 4$ & (2,2) &  (64 ,64 , 64)  &    131072\\
Batch normalization & / & / & / & / &  (64 ,64 , 64) & 128\\
2D transpose convolution & Leaky relu $\alpha = 0.1$ & 32 & $4\times 4$ & (2,2) & (128, 128, 32)  &32768 \\
Batch normalization & / & / & / & / &  (128, 128, 32) & 64\\
2D  convolution & tanh & 1  & $5\times 5$ & (1,1) & (128, 128, 1) &   800 \\
\hline
\end{tabular}
\end{table*}

The goal of the generator is to produce a 128$\times$128 image starting with a vector of 100 elements called the noise vector.
To do so, the generator uses 2D-transposed convolution layers. 
This type of layers preform a similar operation as the traditional convolutional layer, but with the redefinition of the concepts of stride and padding in order to upscale rather than downscale an image \cite{DumoulinConv2016}. 
We used the leaky ReLu activation function with a leak strength of $\alpha = 0.1$ in all the transposed convolutional layers. 
In the final convolutional layer of the generator the $\tanh$ activation was used \cite{radford2015unsupervised}. 
We also included batch normalization layers \cite{Ioffe2015,radford2015unsupervised} for stable GAN training.
Finally, the generator is made of 7242848 trainable parameters (see Table\,\ref{genArchitecture}).

\subsubsection{The physical models}

%Algorithm \ref{GANalgo} lists the need for a data generating distribution $p_{data}(\boldsymbol{x})$. A set of images 

%As training datasets, relevant model images where used. Two types of sources are discussed in this work, hence two types of models are used. These are disk sources and stellar surfaces of stars with a convective outer envelope, such as asymptotic giant branch and red giant stars.
To train the GAN we need to have model images.
In this paper we focus on circumstellar disks of dust and gas as observed in the near-infrared.
We produced model images of circumstellar disks using a radiative transfer code MCMax \cite{Min_2009}.
MCMax was successfully used to model several disks around post-asymptotic giant branch (post-AGB) binaries \cite{Hillen2014,Hillen2015,Kluska_2018}.
It consists of an axisymmetric disk model where the disk scale-height is set by hydrostatic equilibrium.
Our grid of models is based on the best fit model of the disk around IRAS08544-4431 \cite{Kluska_2018}.
We therefore set the central star to have the luminosity and radius of the primary star of IRAS08544-4431 \cite{Kluska_2018}. 
The mass of the central star is $M = 2.4 m_{\odot}$.
We only varied the disk parameters that influence the image in the near-infrared: the inner disk radius $R_{in}$, the power law index describing the surface density as a function of radius for the inner region of the disk $\rho_{in}$ and another such power law index describing the outer region of the disk $\rho_{out}$ (see Table\,\ref{MCMaxParam}).
This gives 455 different models.

Once each model were generated the images were produced at random inclinations ($i$) chosen in a flat distribution of the cosine of the inclinations.
For each model we produced 12 images without the central star half of which in the continuum at $1.5\mu$m and the other half at $2.1\mu$m. 
With this a total of 5\,460 images were generated from MCMax.
These images are sized to $128\times128$ pixels to meet the input size of the discriminator. %These images where originally created at a  $256\times256$ pixels size, but where downsampled to a $128\times128$ pixels size in order to improve computaional efficiecy.
Each time one of these images is sampled to be presented to the discriminator, it is given a random position angle rotation. The distribution of position angles is chosen to be flat.
Upon sampling an image has a 1 in 4 chance to have a uniform 'unresolved' background added. The random value that is added to each pixel value when this occurs follows a uniform distribution between 0 and 0.1 relative to the maximum flux. This was done as certain disk sources are known to have an over-resolved flux which can not be accounted for by radiative transfer models (e.g.\cite{Lazareff2017,Kluska2019,Kluska_2018,Kluska2020}).

Before being fed to the GAN, these images are individually normalized to have pixel values in a range spanning between -1 and 1. The value -1 corresponds to a relative flux of 0 while 1 represents the maximum flux in the image. 
A random zoom on the images from a flat distribution ranging between -10 and 10\% was also applied. Both this zoom and the random rotation of the position angle upon sampling are achieved using bi-linear interpolation.

%The second type of model images for which a GAN was trained are snapshots of 3D radiative-hydrodynamics $CO^5BOLD$ simulations. This code solves the coupled nonlinear equations of  non-local radiative energy transfer and compressible hydrodynamics on a 3D cartesian grid in the presence of a fixed external spherically symmetric gravitational field.  The snapshots where obtained from \citep{Chiavassa_2020} and ?????!!!! , who created these snapshots by using the Optim3D radiative transfer code. Optim3D \citep{Chiavassa_2009} computes the radiative transfer in detail using pre-tabulated extinction coefficients generated with the MARCS code \citep{Gustafsson_2008} and takes into account the Doppler shifts caused by the convective motions.  The extinction coefficients are computed as a function of temperature, density, and wavelength for the solar composition. The temperature, density, and wavelength distribution was optimized to cover the values encountered in the outer layers of the RHD simulations. %Rik: maybe a bit much?

\begin{table*}[t]
\centering
\caption{The parameter range used to construct a grid of MCMax\cite{Min_2009} disk models. }
\label{MCMaxParam}
\begin{tabular}{ c  c  } 
\hline
  Parameter & Grid Range \\
\hline
\hline
$R_{in}$ & [6.0; 6.5; 7.0; 7.5; 8.0; 8.5; 9.0]\\ 
$\rho_{in}$ & \pbox{20cm}{[0.6; 0.3; 0.0; -0.3; -0.6; -0.9;\\  -1.2; -1.5; -1.8; -2.1; -2.4; -2.7; -3.0  ]}  \\
$\rho_{out}$ &  [0.5; 1.0 1.5; 2.0; 3.0.]  \\
\hline
$i$ & flat distribution on $\cos i$, 12 inclinations per model, i between 0\degree and 70\degree \\
PA & flat distribution, randomly rotated each mini-batch \\
Zoom & flat distribution between -10 \% and 10 \%, randomly chosen each mini-batch  
\end{tabular}
\end{table*}

The whole image reconstruction process is divided into two phases A and B.
Phase A is the pre-training of the networks with astrophysical models.
Phase B is the actual image reconstruction phase.

\subsection{Phase A: pre-training}
\label{sec:training}

To be able to perform image reconstructions the GAN needs to go through phase A to be pre-trained on models.
This GAN pre-training is preformed mainly following \cite{goodfellow2014}, including their proposed alteration to the generators gradient. The gradients used for the discriminator where also adjusted.\cite{salimans2016improved}

We quickly recall here the main steps. First the alterations to the gradients of both the generator and discriminator are discussed.

Solving Eq.\,\ref{minmaxProblem} may provide insufficient gradient for the generator to train well \cite{goodfellow2014}. To deal with this problem we maximise $log(D(G(z)))$ for the generator instead of minimizing $log(1 - D(G(z)))$  \cite{goodfellow2014}. 
Doing so results in the same optimum as well as providing a stronger gradient during early training. In deep learning the convention is to implement such optimization problems as minimization problems. 

So with the alteration to the gradient of the discriminator, our optimization problem is reformulated as trying to simultaneously optimize both:
\begin{equation}
     \underset{\boldsymbol{\theta_D}\in \Theta_D }{\operatorname{argmin}} -V(D,G)
    \label{disObjective}
\end{equation}
and
\begin{equation}
     \amn -\mathbb{E}_{\boldsymbol{z}\sim P_{\boldsymbol{z}}}[log(D(G(z)))].
    \label{genObjective}
\end{equation}

The objectives given in equation \ref{disObjective} can be interpreted through the binary cross entropy cost function. This cost function is given by
%\begin{equation*}
%\begin{split}
%         l(\boldsymbol{x}) = -\frac{1}{m}\sum_{i=1}^m & \Big[  y_i\log(D(\boldsymbol{x}_i;\boldsymbol{\theta_D}))   +  \\ &(1-y_i)\log(1-D(\boldsymbol{x}_i;\boldsymbol{\theta_D}))\Big],
%\end{split}
%\end{equation*}

\begin{equation*}
         l(\boldsymbol{x}) = -\frac{1}{m} \sum_{i=1}^m  \Big[  y_i\log(D(\boldsymbol{x}_i;\boldsymbol{\theta_D}))   +  (1-y_i)\log(1-D(\boldsymbol{x}_i;\boldsymbol{\theta_D}))\Big],
\end{equation*}
and is commonly used for binary classification problems. The discriminator can thus be seen to be trained to achieve a target response of $y_i= 1$ for the "real" images and one of $y_i = 0$ for the "fake" images. Meanwhile, the generator is trained to achieve a target response from the discriminator of $y_i = 1$. The generator is thus trained in order to try and convince the discriminator that the images which it creates are a part of the training dataset. 

Based on this formulation of the objectives, \cite{salimans2016improved} proposed an alteration called "one sided label smoothing". When using this technique the ``fake'' images are labelled as 0 whereas the ``true'' ones are labelled as 0.9 instead of 1 when training the discriminator. The target for training the generator remains equal to be 1. Applying one sided label smoothing prevents overconfidence in the discriminator, thereby, making the GAN more likely to converge \cite{Goodfellow2017}.  
The procedure to optimize both \ref{disObjective} and \ref{genObjective}, consists of iteratively updating the component networks and is outlined in appendix \ref{algs} algorithm \ref{GANalgo1} and is describe by hyper-parameters like the number of training iterations, the size of the training dataset and the number of training epochs.

During a training iteration, two mini-batches each consisting of 50 images is sampled.
The first mini-batch consists of a random selection of images from the training dataset. %A random rotation and zoom is applied to these images upon sampling.
The second mini-batch consists of "fake" images.
%is made of a mini-batch of 50 model images and 50 images from the generator randomly ordered.
A ``fake'' image is generated by giving a randomly generated input vector of 100-elements to the generator.
Each element of the input vector is sampled from a Gaussian distribution centered on 0 with a standard deviation of 1.
Both these mini-batches are then used to compute a learning gradient for the discriminator.
After the discriminator is updated, a second mini-batch of 50 noise vectors is sampled. These noise vectors are then used to compute a learning gradient for the generator and update it.

An epoch is a training round where almost the entire set of available images has been presented to the GAN, that is $\lfloor N_{model}/m\rfloor = \lfloor 5460/50 \rfloor = 109$ iterations.
Here we choose to preform the training for 250 epochs.

\begin{figure}[t]
    \centering
    \includegraphics[scale =0.33]{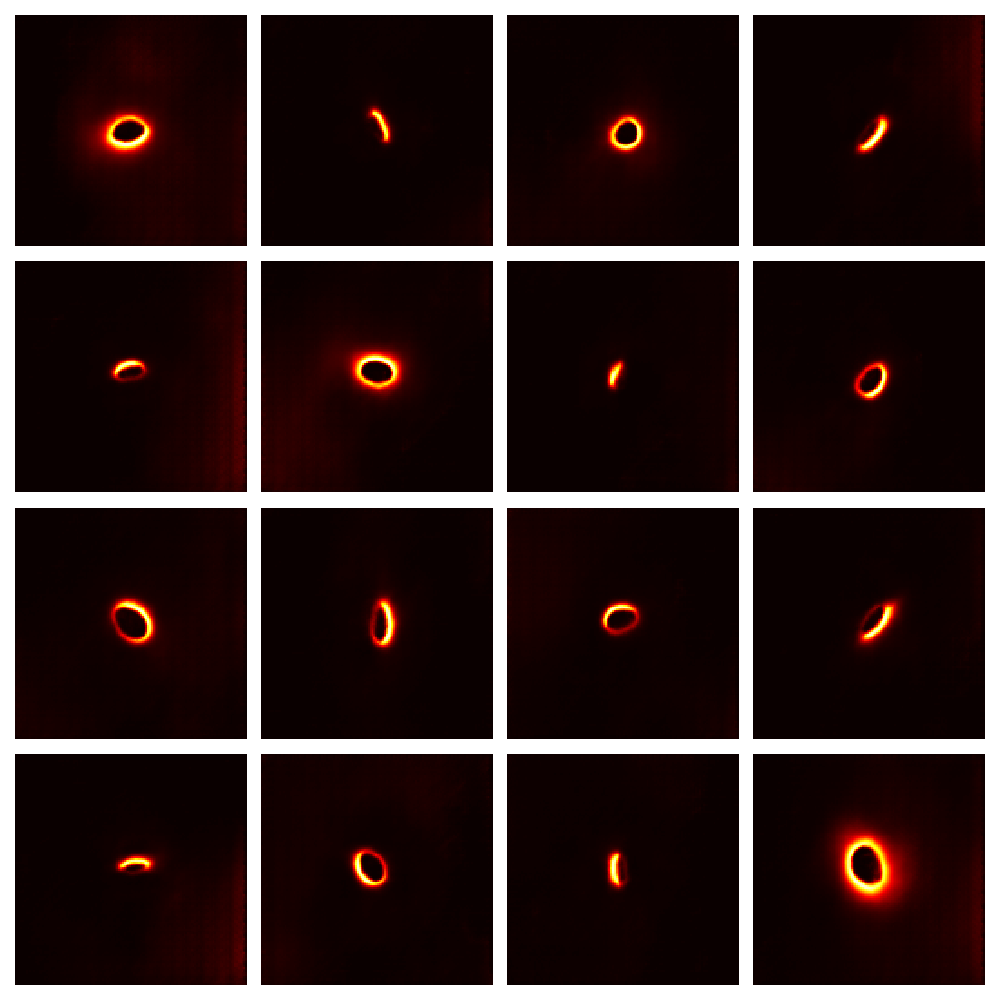}
    \caption{A random selection of disk-images created by the the generator network after GAN-training}
    \label{mosaic}
\end{figure}

We trained the GAN on a Nvidia Tesla P100 GPU, provided by the Flemish SuperComputer. 
GPU training of the GAN takes approximately two hours of physical time. 
A selection of images produced by the generator network after training can be seen on figure \ref{mosaic}.
%\JK{Here more info on the training is necessary with figures on the loss function evolution for ex.} In order to validate that the GAn has reached a stable state both the cost evolution and images created by the generator during training where plotted. The plotted cost evolution displays the mean cost over the iterations which make up an epoch, as a function of epoch. This cost evolution can be seen in figure \ref{}

\subsection{Phase B: image reconstruction}

Once the GAN is pre-trained with MCMax models, it can be used to reconstruct images that reproduce interferometric data. 
In that phase, the generator of the GAN is fine-tuned in order to reconstruct the image taking into account the data likelihood.
To do so, the loss function is adapted to match Eq.\,\ref{eq:Bayes2}.

\subsubsection{Data likelihood}
%The Generator of the GAN produces images with pixel values that range from -1 to 1. A negative flux is of course unphysical.  Therefore, the data likelihood is computed for images, which, are shifted by setting the pixel values to $(x_i+1)/2$, where $x_i$ is the original pixel value.
The data likelihood function is first obtained by calculating the fast Fourier transform of the reconstructed image. The complex visibilities at the relevant spatial frequencies are then obtained by applying a bi-linear interpolation to the Fourier transformed image. 
The squared visibilities and closure phases are computed from these complex visibilities and then compared to the measurements.
The $\chi^2$ for squared visibilities ($S_{j_1,j_2}$) between telescope $j_1$ and $j_2$ is:
\begin{equation}
    \chi^2_{v^2} = \sum_{j_1<j_2} \frac{\big(S_{j_1,j_2}^{data}-S_{j_1,j_2}^{Image}\big)^2}{Var(S_{m,j_1,j_2}^{data})} 
\end{equation}

For circumstellar disks that were observed in the near-infrared the closure phase signal is not showing any phase jumps because of the contribution of the star that is unresolved \cite{Kluska2020}.
It is, therefore, possible to neglect potential phase jumps.
In this case; we can directly use this expression for the closure phases:
\begin{equation}
    \chi^2_{c.p.} = \sum_{j_1<j_2<j_3}\frac{( \beta_{j_1,j_2,j_3}^{data}-\beta_{j_1,j_2,j_3}^{Image})^2}{Var(\beta_{j_1,j_2,j_3}^{data})}
  \label{ch2:liklyhoodclosPh}
\end{equation}
The value then used as the data likelihood is then $f_{data} = (\chi^2_{v^2} + \chi^2_{c.p.})/N_{observables}$ where $N_{observables}$ is the sum of the number of $V^2$ and CPs.
%The computation of this data likelihood term has been implemented using Keras-tensorflow backend functions, allowing for the computation of  $\Delta_{\boldsymbol{\theta}_G}  f_{data}(G(\boldsymbol{n},\boldsymbol{\Theta}_G))$ by the Keras framework. This allows for Eq.\, \ref{objectiveImageRec} to be minimized with respect to the reconstructed image through the use of a gradient based learning rule. 

We implemented the SPARCO approach \cite{Kluska_2014} that consists in adding the contributions of a central star or stars to the reconstructed image. 
%In the SPARCO framework the geometry of the system is modeled using simple geometrical contributions and the different chromaticity of the different components can be accounted for using power law approximations. 
The functionality to add both a uniform disk source and a point source is implemented in our code \cite{Hillen2016A&A}. 
This allows for the contributions of the central stars to be removed from the image and provide an image of the circumstellar environment with enhanced quality.
It also allows to use the chromatic data from all the channels at once and reconstruct a single image that is valid at all sampled wavelengths.

\subsubsection{Regularisation}

The regularisation consists of the negative logarithm of the discriminators output for the reconstructed image.
The objective function for which the generator is optimized is expressed as follows:
\begin{equation}
    f = f_{data}(G(\boldsymbol{z},\boldsymbol{\Theta}_G))-\mu log(D(G(\boldsymbol{z},\boldsymbol{\Theta}_G));\boldsymbol{\Theta}_D)
    \label{objectiveImageRec}.
\end{equation}
The first term is the data-likelihood, which computes the agreement between the observations and the image produced by the generator $G(\boldsymbol{n},\boldsymbol{\theta}_G)$ and the observations. The idea here is that this regularization will constrain the generator to produce image which remain close to $P_{model}(\boldsymbol{x})$.

\subsubsection{Mitigating network imperfections}

When simply optimizing the objective given in Eq.\, \ref{objectiveImageRec} for a random  $\boldsymbol{z}$ two problems occurred.

\begin{figure}
    \centering
    \includegraphics[scale=0.6]{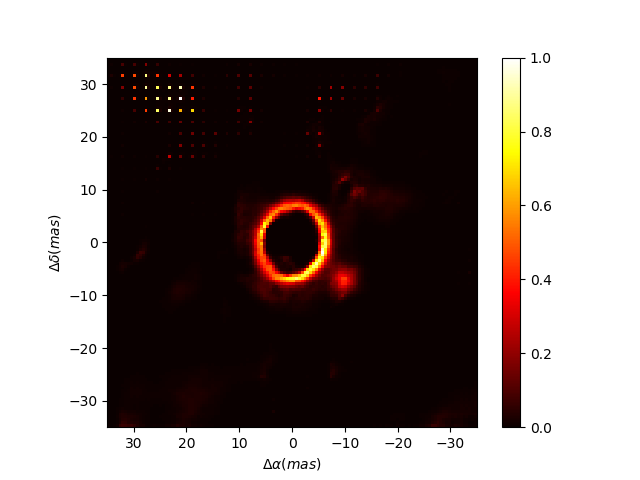}
    \caption{Example of checkerboard artifacts appearing when optimizing \ref{objectiveImageRec}. This example occured after 500 epochs, with a value of $\mu=2$.}
    \label{fig:checkerboard}
\end{figure}

\paragraph{checkerboard artifacts:}
The first one is the creation of checkerboard artifacts in the reconstructed image (an example of such pattern is presented Fig.\, \ref{fig:checkerboard}).
These types of artifacts appear more often after a large number of gradient descent steps of fine-tuning the generator. For larger values of $\mu$ they appear earlier.  These gradient descent steps are referred to as epochs. These types of artifacts are typical for two processes: the forward-propagation of up-sampling
layers and the back propagation of convolutional layers used to compute the gradient \cite{sugawara_shiota_kiya_2019}. The first is present in the Generator and the second is introduced by the use of the discriminator in equation \ref{objectiveImageRec}. These types of artifacts are clear deviations of the prior which we wish to impose.
For values of $\mu$ lower than 10, the optimization of data likelihood has, however, largely obtained before these features become prominent.  
%From experience such problem occurs after more than 300 gradient descent steps. 
We have therefore chosen to fine tune the generator for 250 epochs long, thereby, avoiding these artifacts. This lower amount of epochs is also beneficial for the run time. As a consequence, Eq.\, \ref{objectiveImageRec} will never be fully minimized. A different amount of epochs, may yet improve results further. Alterations to the neural network architecture intended to avoid these types of artifacts (e.g. \cite{sugawara_shiota_kiya_2019,gao2017pixel}) were not explored and are beyond the scope of this contribution.

\paragraph{Image variability}
The second problem is that the images are not consistent between runs, as they have different noise vectors. 
This is likely a consequence of both the optimization being stopped early and the regularization not being convex, causing the training procedure to get stuck in different local minimum, depending on the start position of the generator set by the input noise vector. In order to obtain a near constant image is was chosen to restart the image reconstruction 200 times and use the median image as the final reconstructed image.

\subsubsection{Actual image reconstruction process}
\label{sec:phaseB}
These alterations bring us to the image reconstruction algorithm that is actually used.
%The image reconstruction process is made by training the generator ($G$) only (i.e., the discriminator $D$ weights are kept fixed) to produce images that both reproduce the data and obtain a good score from the discriminator. To do so a double iterative loop is designed.
The procedure consists of a double iterative loop.
The inner loop, iterating over a number of epochs, is made of learning iterations with a constant input vector. During this inner loop the generator network is updated to descend the gradient of Eq.\, \ref{objectiveImageRec} for a fixed input $\boldsymbol{z}$.
Once the learning have reached a certain number of iterations (called $n_\mathrm{epoch}$) the final image is stored and we start another iteration in the general loop.
When restarting a new input $\boldsymbol{z}$ is randomly chosen from the noise prior $P_z(\boldsymbol{z})$ and the generator is reset to its state from the end of the pre-training phase.

% The stored images from each restart are first shifted the pixel value range of the generator to a range between 0 and 1. After this they are normalized to have a total flux equal to 1, thereby ensuring these images occupy the same Fourier space. 
The median of these images is then adopted as our final image reconstruction. This procedure is more formally described by the pseudo-code given in appendix \ref{algs} algorithm \ref{GANalgo2}.

%In the second term $\mu$ is the hyperparameter which tunes the strength of the regularization term. 
%The function $f_{prior}$ is chosen to be $f_{prior} = -\log(x)$ , this way the regularization term matches both the expected form based on Bayesian considerations and has the same behaviour as the binary-cross entropy loss function when the labels are always set to 1. 
%This can be interpreted as requiring that the discriminator should classify the reconstructed image as part of the distribution $ P_{prior}(\boldsymbol{x})$. 

%The input which supplied to the generator is kept constant during this training after initially being sampled from the noise distribution $P_{\boldsymbol{z}}(\boldsymbol{z})$. By doing so the fine-tuning of the generator solves a similar problem as defined by the traditional image reconstruction objective function. 

\section{Validation on artificial datasets}\label{validation}

We created three artificial datasets to validate our method.
To keep it realistic we have used a (u, v)-coverage of existing data from the PIONIER instrument of the VLTI. 
We added Gaussian noise to the $V^2$ and CPs.
These real datasets are observations of IRAS08544-4431 from \cite{Hillen2016A&A} and HD45677 from \cite{Lazareff2017,Kluska2020}.
The corresponding (u, v)-coverages can be found in Appendix \ref{UVcov}.
Hereafter, these datasets will be referred to as dataset A, B and C respectively. The (u, v)- coverages and noise of datasets A and C are based on the observations of IRAS08544-4431, while dataset B is based on HD45677.

The images used for datasets A and B are taken from the grid of models used to train the GAN. The images used for all datasets have sizes of $128\times128$ pixels images with a field of view of 70\,mas. Dataset B is further discussed in the appendix \ref{datasetB}. For dataset C an image consisting of a model image from the same grid with an added Gaussian flux. This Gaussian has an amplitude of 0.5 normalized flux (relative to the maximum flux in the image), a standard deviation of $1.8 mas$ in both the directions on the sky. The Gaussian is centered at $\Delta \alpha = 10.8 mas$ and $\Delta\delta = 8.4 mas$.  This element was included in order to test the capability of the framework in reconstructing unexpected features.

Since the model images do not contain central stars, the contribution of a central star was added using SPARCO with values to make the datasets realistic.
Hence, the stellar-to-total flux ratio of the primary at 1.65$\mu$m is $f_{prim}= 0.597$,  the diameter of the primary is $D=  0.5 mas$ and the spectral index for the circumstellar environment $d_{env}=0.42$. 
The datasets created in this way can be found in appendix \ref{artificialData}.

%In order to validate that the often used L-curve method yields an optimal value for the hyperparameter in this reconstruction method, this method was applied to the artificial datasets. 
To validate the quality of the reconstructed images we used two metrics for comparing the used model image and the reconstructed one. 
We used the mean squared error (MSE) metric \cite{Renard2011} and normalized cross correlation (NCC) \cite{M87}. 
When we apply these metrics we first normalize the images such that the total flux in both images equals unity, ensuring these images occupy the same Fourier space.

The normalized cross-correlation is given by:
\begin{equation}
    \rho(\boldsymbol{X,Y}) = \frac{1}{N}\sum_i\frac{(X_i-\langle X \rangle)(Y_i-\langle Y \rangle)}{\sigma_X \sigma_Y}
\end{equation}
Here the summation is over all the $N$ pixels of the model image $X$ and the reconstructed image $Y$. $\langle X\rangle$ and $\langle Y\rangle$ are the mean pixel values of the images. $\sigma_X$ and $\sigma_Y$ are the standard deviations of the pixels in the two images. The normalized cross-correlation quantifies the similarity between images. A value of -1 implies perfect anti-correlation between images, 0 implies no correlation, and 1 implies perfect correlation.

%The L-curve containing the tested values of $\mu$, can be seen in appendix \ref{L-curves}. The L-curve behaviour is atypical as $f_{prior}$ appears to stagnate for the lowest values of $\mu$. This may be a consequence of the lower number of iterations chosen to fine-tune the generator in itself may prevent an over optimization of the data-likelihood, thereby preventing the kernels of the generator to change too much from it's initial state. For the highest values of $\mu$ a strong decrease in $f_{prior}$ still occurs. Despite this different behaviour compared to L-curves with conventional regularization's, the optimal image for both datasets A and B appears to fall at the corner between a data dominated and regularization dominated regime. For dataset C the optimal value of $mu$ falls in the data dominated regime.

%The increase in the regularization value and lower values of $\mu$ appears to stagnate. The reason for this may be two fold. The lower number of iterations chosen to fine-tune the generator in itself may prevent an over optimization of the data-likelihood, thereby preventing the kernels of the generator to change too much from those obtained during the GAN training. Another explanation could be that the resulting image will always score relatively well in the discriminator as the data is based on one of the models on which the GAN was trained.

\begin{figure*}[t!]
    \begin{subfigure}{0.49\linewidth}
    \includegraphics[width=\textwidth]{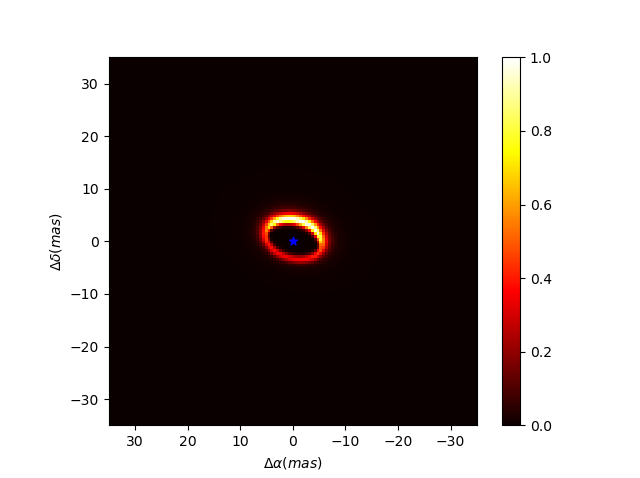}%this is image 2395 in the directory listing
    \end{subfigure}
    \qquad
    \begin{subfigure}{0.49\linewidth}
    \includegraphics[width=\textwidth]{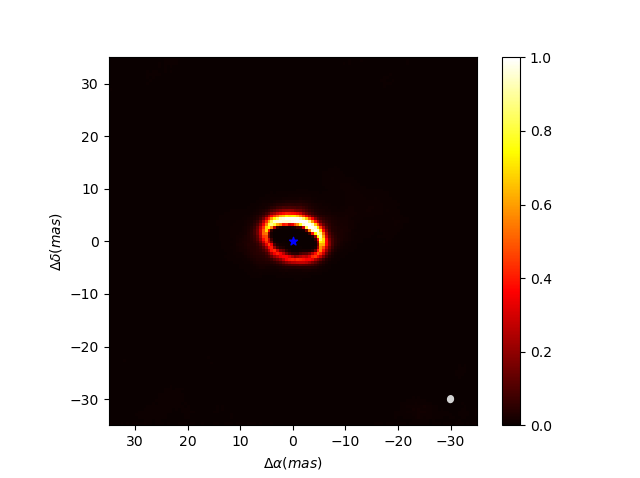} 
    %this is image 3395 in the directory listing
    \end{subfigure}
    \caption{Right: The model image used for constructing the artificial dataset. The used model image has  $R_{in} = 7.5au$ $\rho_{in} = -0.6$ $\rho_{out} = 1.0$ $i =45.3 \deg$ and $\lambda = 1.5\mu m$. Left: The image reconstruction using $\mu =5$. The location of the added central star is indicated by the blue star. The gray ellipse on the reconstruction displays the beamsize. The beam size is defined as twice the Gaussian FWHM fitted to the interferometric point spread function.}
    \label{ModelImagesArtificialA}
\end{figure*}

The comparison between the true images and reconstructions of dataset A, B and C can be found in appendix \ref{tables}. 
For dataset A the reconstruction with the optimal MSE and NCC with respect to the model is that with $\mu = 5$, this reconstruction can be seen on figure\ref{ModelImagesArtificialA} next to the model used to create the corresponding artificial dataset. Figure\ref{ModelImagesArtificialA} it can be seen that our image reconstruction scheme is capable of reproducing the feature seen in the model rather well. A sharp inner rim can be seen and the flux away from the inner most part of the disk appears flat. 
%Both of these values are in the data-likelihood dominated regime of the L-curve, rather than the bend of the curve.
The results on dataset B are discussed in Appendix \ref{datasetB}. 
Figure \ref{ArtificialObservablesC} displays the image used to construct dataset C next to the optimal reconstruction.  Both the disk and the Gaussian flux are recovered well using our method. The recovery of the Gaussian illustrates that the method is capable of recovering unexpected features. The prior thus imposed using our framework, appears to be soft enough allowing for the complexity of a source to be recovered.

%The comparison between the model images and reconstructions of dataset A can be found in table \ref{ModelComparison}. The reconstruction with the optimal MSE and SSIM with respect to the model is that with $\mu = 2$.

\begin{figure*}[t]
    \begin{subfigure}{0.49\linewidth}
    \includegraphics[width=\textwidth]{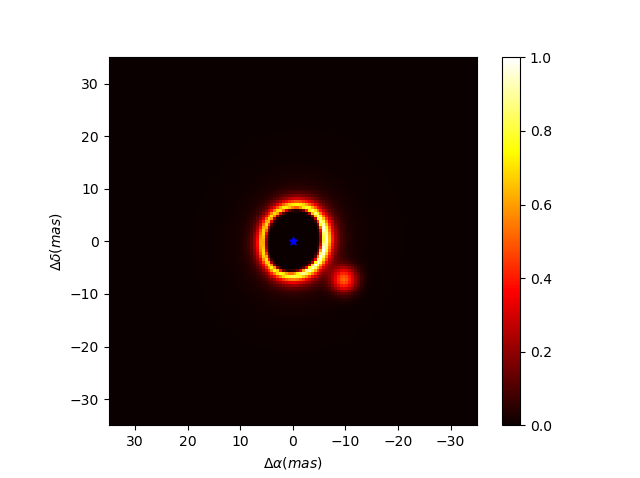}
    \end{subfigure}
    \qquad
    \begin{subfigure}{0.49\linewidth}
    \includegraphics[width=\textwidth]{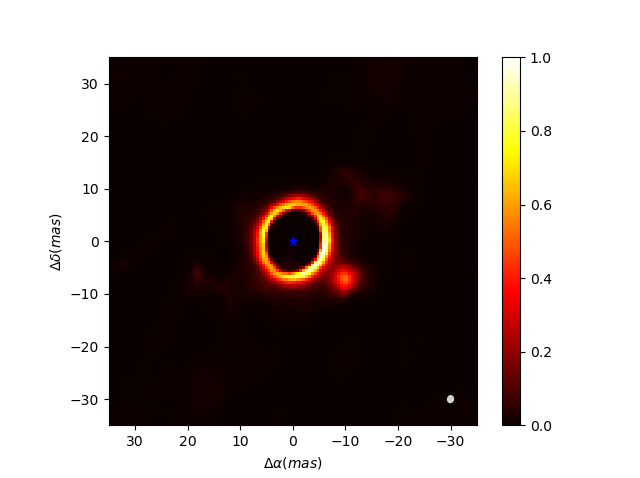}
    \end{subfigure}
    \caption{Left: the model disk image with added Gaussian used to construct artificial dataset C . The used model image has  $R_{in} = 9au$ $\rho_{in} = -0.6$ $\rho_{out} = 0.5$ $i =25 \deg$ and $\lambda = 1.5\mu m$. Right: The image reconstruction with $\mu = 2$ using our method. The blue star marks the position of the added central star in both images. The colorbar indicates the flux normalized to the maximum in the image. The gray ellipse on the reconstruction displays the beamsize.}
    \label{ArtificialObservablesC}
\end{figure*}
\section{Conclusion}\label{conclusion}
In this work the use of a GAN to reconstruct images from optical interferometric data was explored.
In this framework a GAN is first trained on a set of model images representing the prior distribution. The component networks could then be used to preform image reconstructions, by fine-tuning the generator and using the discriminator as part of the regularization.

This framework was applied to realistic, artificial datasets based on model images which were a part of the training dataset of the GAN. 
%It was shown that for these artificial datasets, the framework provides reconstructions which more closely resemble the images on which the artificial dataset is based, compared with conventional image reconstruction methods.  
It was shown that for these artificial datasets, the framework provides reconstructions of high quality which are very similar to the original images.
The images reconstructed for these datasets are extremely convincing and appear as almost axis symmetric disks. Artifacts which typically appear away from the inner rim of the disk did not appear in our image reconstructions. 

The method appears not only capable of doing this for data based on the models used in training, but also when an unexpected feature is introduced in the data. This is illustrated by the reconstruction on dataset C. The Gaussian addition to this data-set is reproduced. This result indicates that the reconstruction method appears capable of allowing unexpected features, which are present in the data, to appear in the image.

To the best of our knowledge, this is the first time that the use of neural networks was explored in order to implement a Bayesian prior for the image reconstruction of interferometric data.
For a first exploration of such a methodology, the results obtained here, appear promising.

%There seem to be a large amount of improvements possible to the image reconstruction scheme presented here. 
There are still improvements that we plan to add to the image reconstruction scheme presented here.
So far the main limitation is the computational cost of the method. Currently it takes about 13 minutes on a NVIDEA Tesla P100-SXM2-16GB gpu in order to reconstruct a single image with the framework presented here. This proves to be prohibitive with regards to bootstrapping the data in order to asses the significance of the features visible in the image.

The main contribution to this large computational expense in the reconstruction is the size of the generator network which needs to be retrained. This size is necessitated by the training of the GAN itself, as a generator with insufficient capacity will be incapable of mapping the input noise vectors to images of the distribution $P_{data}$. Exchanging the generator, for a different network capable of producing images does not give satisfactory results, as the discriminator appears to constrain the generator, rather than provide a gradient towards $P_{data}$. A network capable of providing such a gradient would thus drastically improve the computational cost of such reconstructions, and could potentially even be implemented in traditional image reconstruction algorithms.

%A minor improvement could be made by training the GAN on Fourier transformed disk images rather than the model images themselves. This would allow for a more direct comparison between the reconstruction, which is performed is Fourier space and the observables. This makes the computation of the data-likelihood and it's corresponding gradients less complex. In order to do so, the GAN would also need to be adjusted as to produce 2 output images, representing both the real and complex part of the Fourier transform. This can be achieved by setting the number of kernels in the final layer of the generator network to two.

%Beyond this, a variety of network architectures can be tested in the same framework, some of which may prove more resilient to the to the pixel like artifact which emerge when the regularisation term is over optimized. 

Other schemes to use neural networks to reconstruct image can also be considered. One method that may be of interest is the fitting the input of a generative model such as a GAN's generator or a variational auto-encoder\cite{bora2017}. This results in images which follow the models extremely closely. Additional freedom can be introduced  by including a sparse addition to the image\cite{Dhar2018}. 
%There also is no good theoretical basis explain why this method is capable of creating images which more closely resemble the training dataset. Despite this, results appear convincing. \RC{uncertain about this}

Both the speed at which new developments are made in deep learning, and the ever increasing amount of computation power available make it seems likely that deep-learning regularization's will be further developed and will very likely have an important role to play in interferometric imaging.

\acknowledgments
%J>K
J.K. acknowledges support from the research council of the KU Leuven under grant number C14/17/082 and from FWO under the senior postodctoral fellowship (1281121N).
VSC

\appendix

\section{Used neural network layers}\label{app:layers}
When training a GAN two neural networks need to be defined a discriminator network and a generator network. These networks where constructed using fully connected layers, convolutional layers and batch normalization layers. Here a short description of these types of layers is given.

\paragraph{Fully connected layers} used in our networks the pre-activation outputs are computed using either
\begin{equation}
    o_j = \sum_i a_{ij}\cdot x_i + b_{ij} \qquad  \text{or} \qquad o_j = \sum_i a_{ij}\cdot x_i
\end{equation}
depending if one uses biases ($b_ij$) or not. Both $\{a_{ij}\}$ and $\{b_{ij}\}$ are trainable parameters which are optimized using a gradient based learing rule\cite{Goodfellow-et-al-2016}.

\paragraph{Convolutional layers} are based on  discrete convolution operations\cite{DumoulinConv2016,Goodfellow-et-al-2016}.
a chosen number of kernels of a chosen size are used to perform these operations. It are the values in these kernels which constitute the trainable parameters of a convolutional layer.
The pixel values of a pre-activation output of a convolution with such a kernel is computed as follows:
 \begin{equation}
     \boldsymbol{S}(i,j,k)  = \sum_m\sum_n\sum_l \boldsymbol{I}(i-m,j-n,l)\boldsymbol{K}(m,n,l,k)
     \label{ch3:ConvolutionLayerEq},
 \end{equation}
where $\boldsymbol{I}$ represent input into the layer, this can be a grayscale image with a size equal to 1 in the direction of $k$, an ''rgb'' image with a size equal to 3 in the direction of $k$ or the output of a previous convolutional layer which has a depth equal to the number of kernels used in the previous layer.  

$\boldsymbol{K}$ is the convolutional kernel and $\boldsymbol{S}$ the output corresponding to the kernel. 
The indexes $m$ and $n$ are summed over the kernel size.
In a convolutional layer a ''stride'' is also defined.
The stride defines for which values of $i$ and $j$ an output is computed. An output is computed for every value of $i$ and $j$ which is a multiple of the chosen stride in the relevant direction. A stride larger than 1 results in a downsized output image $\boldsymbol{S}$. When a stride equal to two is used the number of pixels in $\boldsymbol{S}$ will be half that of the input images $\boldsymbol{I}$.

%The stride defines if the convolution operation is made for every pixel (stride = 1$\times$1) or skip one (stride = 1$\times$1) or more pixels. For some layers we have used a stride of 2$\times$2 pixels meaning that we perform the convolution every two pixels. 

Finally, the outputs of the both types of layers are further processed by activation functions, which introduce non-linearity into the network, allowing it to tackle more complex tasks.

\paragraph{Batch normalization} works by first normalizing the individual images produced by a convolutional layer by using:
\begin{equation}
    \hat{x}_i = \frac{x_i - \overline{x}_B}{\sqrt{\sigma^2_B + \epsilon}}
\end{equation}
Here $\overline{x}_B$ and $\sigma_B^2$ are the average and variance of the output images $x_i$ of the previous layer during training on one mini-batch of data and  $\epsilon = 0.001$ is a constant added to $\sigma^2_B$ for numerical stability.  These normalized outputs are then shifted and re-scaled using:
\begin{equation}
    y_i= \gamma\hat{x}_i+ \beta \equiv BN_{\gamma,\beta}(x_i)
    \label{ch3:batchnorm}
\end{equation}
Where $\gamma$ and $\beta$ are trainable parameters which are optimized by the used gradient based learning rule. 
In our case, batch normalization layers where applied before the activations of the previously listed convolutional layers. 
Two additional trainable parameters are thus added per output image of each convolutional layer as can be seen in table \ref{genArchitecture}.
During inference $\overline{x}_B$ and $\sigma_B^2$ are replaced by a moving average $\overline{x}_M$ and moving variance $\sigma_M^2$. The values of $\overline{x}_M$ and $\sigma_M^2$ are updated each training iteration using
\begin{equation}
    \overline{x}_M = \overline{x}_M\cdot \alpha + \overline{x}_B\cdot (1-\alpha)
\end{equation}
and 
\begin{equation}
    \sigma_M^2 = \sigma_M^2\cdot \alpha + \sigma_B^2\cdot (1-\alpha).
\end{equation} Here $\alpha = 0.99$ was used.

\section{The algorithms}\label{algs}
The image reconstruction procedure presented in this paper consists of two phases. The first preparatory phase consists of training a GAN. A pseudocode detailing this training procedure is outlined in algorithm \ref{GANalgo1}. After a GAN is trained, images can be reconstructed using both the component networks of the GAN. The procedure used to reconstruct images can be found in algorithm \ref{GANalgo2}. 
\begin{algorithm}[h!]
\textbf{Require:}
  generator network G with parameters $\theta_G$\;
  discriminator network D with parameters $\theta_D$\;
  A chosen number of training iterations\;
  %The hyperparameter $k$\;
  A chosen mini-batch size $m$\;
  data generating distribution $p_{data}(\boldsymbol{x})$\;
  noise prior distribution $p_{data}(\boldsymbol{x})$\;
  A gradient-based learning rule\;
  \hrulefill
  
  \For{number of training iterations}{
  %\For{k steps}{
   $\bullet$ Sample mini-batch of $m$ noise samples $\{ \boldsymbol{z}^{(1)},...,\boldsymbol{z}^{(m)}\}$ from noise prior $p_g(\boldsymbol{z})$
   
   $\bullet$ Generate a mini-batch of $m$ examples $\{ \boldsymbol{x}^{(1)},...,\boldsymbol{x}^{(m)}\}$, where $\boldsymbol{x}^{(i)} = G(\boldsymbol{z}^{(i)};\boldsymbol{\theta_G})$
   and provide corresponding output labels $y^{(i)}$ = 0
   
   $\bullet$     Sample minibatch of $m$ examples $\{ \boldsymbol{x}^{(m)},...,\boldsymbol{x}^{(2m)}\}$ from data generating distribution $p_{data}(\boldsymbol{x})$ and provide corresponding output labels $y^{(i)}$ = 0.9
   
   $\bullet$ Update the discriminator by using the chosen learning rule to descending its stochastic gradients:
    \begin{equation*}
    \begin{split}
        -\Delta_{\boldsymbol{\theta_D}}\frac{1}{2m}\sum_{i=1}^{2m} & \Big[  y^{(i)}\log(D(\boldsymbol{x}^{(i)};\boldsymbol{\theta_D}))   + \\  & (1-y^{(i)})\log(1-D(\boldsymbol{x}^{(i)};\boldsymbol{\theta_D}))\Big] \\
    \end{split}
    \end{equation*}
    
    %\textbf{end for}
   %}
   {
   $\bullet$ Sample mini-batch of $m$ noise samples $\{ \boldsymbol{z}^{(1)},...,\boldsymbol{z}^{(m)}\}$ from noise prior $p_g(\boldsymbol{z})$ and provide corresponding output labels $y_i$ = 1
   
   $\bullet$ Update the generator by using the chosen learning rule to descending its stochastic gradients:
   \begin{equation*}
       -\Delta_{\boldsymbol{\theta_G}}\frac{1}{m}\sum_{i=1}^m y^{(i)}\log(D(G(\boldsymbol{z};\boldsymbol{\theta_G});\boldsymbol{\theta_D}))
   \end{equation*}
   
   \textbf{end for}
  }
 }
\caption{1 Mini-batch stochastic gradient descent training of generative adversarial nets. The gradient-based updates can use any standard gradient-based learning rule.}
\label{GANalgo1}
\end{algorithm}

\begin{algorithm}[h]
\textbf{Require:}
  generator network G with parameters $\theta_G$ determined by the GAN training\;
  discriminator network D with parameters $\theta_D$ determined by the GAN training \;
  Achosen number of times to reset the generator and resample a noisevector\;
  A chosen number of epochs for fine tuning the generator\;
  The hyperparameter $\mu$\;
  noise prior distribution $p_{g}(\boldsymbol{z})$\;
  A gradient-based learning rule\;
  \hrulefill
  
  \For{number of resets}{
  $\bullet$ set the generator parameters to those obtained at the end of GAN training\;
  $\bullet$ Sample a noise samples $\boldsymbol{z}$ from noise prior $p_g(\boldsymbol{z})$.\;
  \For{epochs}{
   $\bullet$ Update the generator by using the chosen learning rule to descend the gradient:
    \begin{equation*}
    \begin{split}
        \Delta_{\boldsymbol{\theta_G}}f = &  f_{data}(G(\boldsymbol{z},\boldsymbol{\theta}_G))\\ & -\mu \log(D(G(\boldsymbol{z},\boldsymbol{\theta}_G));\boldsymbol{\theta}_D)
    \end{split}
    \end{equation*}
    \textbf{end for}
   }{
   $\bullet$ Shift the image $G(\boldsymbol{z},\boldsymbol{\theta}_G))$ to have a flux range between 0 and 1. \;
   $\bullet$ Normalize the image $G(\boldsymbol{z},\boldsymbol{\theta}_G))$ to have a total flux equal to 1. \;
   $\bullet$ store the normalized image of this iteration for\;
   \textbf{end for}\;
   $\bullet$ compute the median image out of the set of normalized images\;
  }
 }
\caption{ORGANIC image reconstruction procedure.}
\label{GANalgo2}
\end{algorithm}

\section{Dataset B}\label{datasetB}
\begin{figure*}[h]
    \begin{subfigure}{0.49\linewidth}
    \includegraphics[width=\textwidth]{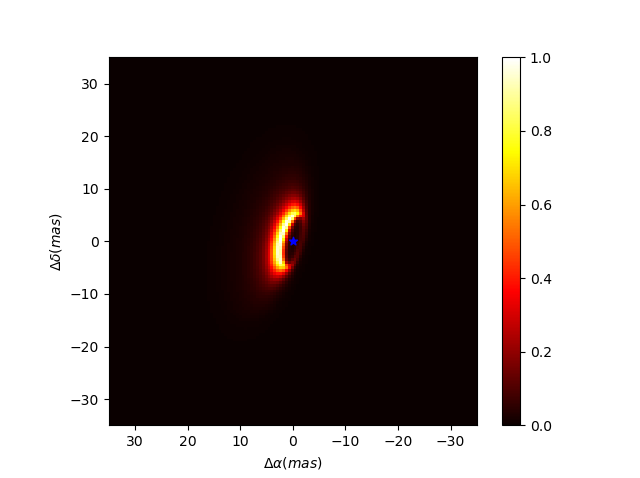}
    \end{subfigure}
    \qquad
    \begin{subfigure}{0.49\linewidth}
    \includegraphics[width=\textwidth]{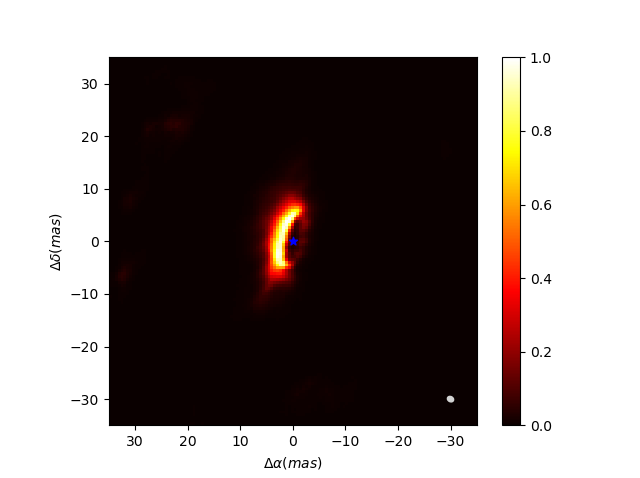}
    \end{subfigure}
    \caption{Right: The model image used for constructing dataset B. The used model image has  $R_{in} = 7.0au$ $\rho_{in} = -2.4$ $\rho_{out} = 1.5$ $i =63.7 \deg$ and $\lambda = 2.1\mu m$. Left: The image reconstruction using $\mu =5$. The location of the added central star is indicated by the blue star. The gray ellipse on the reconstruction displays the beamsize.}
    \label{ArtificialObservablesB}
\end{figure*}
As mentioned in the main text an artificial dataset was created based on the baselines of observations of HD45677 previously used in \cite{Kluska2020}. As with the dataset presented in the main text, an L-curve was plotted for various reconstructions and these reconstructions where compared to the models on which the data was based. %The L-curve plotted for this dataset can be seen in appedix \ref{L-curves} on figure \ref{LcurvArtHD}.
The MSE and NCC for the preformed reconstructions can be found in appendix \ref{tables}. The image reconstruction with $\mu =1$ can be seen on figure \ref{ArtificialObservablesB}. This reconstruction is not as convincing as those mentioned in the main text. The most likely cause for this is the spacer UV-coverage, as this is an important factors in determining the quality of a reconstruction \cite{Renard2011}.

\section{About ADAM optimization}
The chosen gradient based learning rule is the Adam optimization algorithm \cite{kingma2014adam}. 
During phase A the learning rate was set to $\alpha = 0.0002$ and the first moment’s exponential decay rate to $\beta_1 = 0.5$ \cite{radford2015unsupervised}.   During phase B the learning rate was set to $\alpha = 0.0002$ and the first moment’s exponential decay rate to $\beta_1 = 0.91$
In both cases the second moment exponential decay rate and the tolerance parameter we used $\beta_2= 0.999$ and $\mu_\mathrm{ADAM} = 10^{-7}$ respectively.

\section{Comparison between true images and reconstruction}\label{tables}
\begin{table*}[h]
\centering
\caption{The MSE and NCC between the model and reconstructions of dataset A using different values of $\mu$. The optimal values of the MSE and NCC are indicated in bold. The data-likelihood term of the final image $f_{data}$, its components per type of observable (over the number of that observable type.) and the "regularization" term $-\log(D(x_f))$ are also listed.}
\label{ModelComparison}
\begin{tabular}{ c  c c c c c c} 
\hline
\hline
$\mu$  & MSE  & NCC & $\chi^2_{v^2}/N_{v^2}$ & $\chi^2_{c.p.}/N_{c.p.}$ & $f_{data}(x_f)$& $-\log(D(x_f))$\\
\hline
$10^{-6}$ & $7.4979\cdot 10^{-9}$ & 0.9798 & 1.213  &   1.099  &    1.170 & 0.8115067 \\
0.0001 & $7.4038 \cdot 10^{-9}$ & 0.9801 &  1.244  &  1.087 &   1.185& 0.8115067  \\
0.001 & $7.9423 \cdot 10^{-9}$ & 0.9786 &   1.218  &  1.104    &    1.174& 0.8115043   \\
0.01 & $7.3051 \cdot 10^{-9}$ & 0.9803 &   1.215  &  1.102    &    1.172&  0.8115103 \\
0.1 & $7.0300 \cdot 10^{-9}$ & 0.9811 &  1.226   &   1.099   &    1.178&  0.8115058   \\
1 & $4.1973\cdot 10^{-9}$ &  0.9889 &   1.138  &   1.225  &     1.172& 0.8114951 \\
2 & $2.9382\cdot 10^{-9}$ &  0.9924 &  1.084   &  1.316    &   1.172& 0.8114925  \\
5 & $\mathbf{2.7834\cdot 10^{-9}}$ &  \textbf{0.9930} &  1.043  &  1.411    &   1.182& 0.8114937  \\
10 & $3.3611\cdot 10^{-9}$ & 0.9913 &  1.039   &   1.469   &    1.202&  0.8114935  \\
50 & $9.1718\cdot 10^{-9}$ & 0.9753 &  1.558   &   1.221   &    1.431&  0.8114618  \\
100 & $1.503\cdot 10^{-8}$ &  0.9592 &  1.593  &   2.390   &    1.895&  0.8114126 \\
\hline
\hline
\end{tabular}
\end{table*}
\vspace{3cm}

%\begin{table}[h!]
%\centering
%\caption{The MSE and  between the model and reconstructions of dataset C using different values of $\mu$. }
%\label{ModelComparisonC}
%\begin{tabular}{ c  c  c c c c } 
%\hline
%\hline
%$\mu$  & MSE & NCC & $\chi^2_{v^2}/N_{v^2}$ & $\chi^2_{c.p.}/N_{c.p.}$ & $f_{data}$\\
%\hline
%$10^{-6}$  & $1.9070\cdot 10^{-9}$&   0.9884 &  0.977  &   0.970   &   0.974    \\
%0.0001  & $2.0357\cdot 10^{-9}$&  0.9874 & 0.980    &   0.993   &    0.985  \\
%0.001  & $2.0014\cdot 10^{-9}$&  0.9877 & 0.981   &   0.984  &   0.982   \\
%0.01  & $1.875\cdot 10^{-9}$&  0.9885 & 0.976   &   0.981  &    0.978   \\
%0.1  & $1.9051\cdot 10^{-9}$& 0.9886 &  0.983   &   1.011  &   0.994   \\
%1  & $1.4291\cdot 10^{-9}$&  0.9925 &   0.984  &   0.945  &    0.969   \\
%2  & $1.2118\cdot 10^{-9}$ & 0.9937 &  0.984   &   0.941   &   0.968    \\
%5  & $1.4061\cdot 10^{-9}$&  0.9927 &   1.012  &   0.921   &    0.976   \\
%10  & $1.4952\cdot 10^{-9}$&  0.9908 &   1.034 &   0.936   &    0.995  \\
%50  & $1.9156\cdot 10^{-8}$&  0.8642 &   3.763  &   6.275   & 4.767      \\
%100 &$4.0257\cdot 10^{-8}$& 0.6792  &   7.768  &   11.512   &  9.265   \\
%\hline
%\hline
%\end{tabular}
%\end{table}
\begin{table}[h!]
\centering
\caption{The MSE and  between the model and reconstructions of dataset B using different values of $\mu$. The optimal values of the MSE and NCC are indicated in bold. The data-likelihood term of the final image $f_{data}$, its components per type of observable (over the number of that observable type.) and the "regularization" term $-\log(D(x_f))$ are also listed.}
\label{ModelComparisonB}
\begin{tabular}{ c  c  c c c c c} 
\hline
\hline
$\mu$  & MSE & NCC & $\chi^2_{v^2}/N_{v^2}$ & $\chi^2_{c.p.}/N_{c.p.}$ & $f_{data}$& $-\log(D(x_f))$\\
\hline
$10^{-6}$  & $6.2619\cdot 10^{-9}$&   0.9774 &  3.301  &   0.927   &   2.398&  0.8113948  \\
0.0001  & $6.0988\cdot 10^{-9}$&  0.9780 & 3.429    &   0.937   &    2.481& 0.8113982 \\
0.001  & $6.2362\cdot 10^{-9}$&  0.9775 & 3.428   &   0.925   &    2.476 & 0.8113952 \\
0.01  & $6.3580\cdot 10^{-9}$&  0.9719 & 3.737    &   0.949   &    2.676 & 0.8113961 \\
0.1  & $6.2111\cdot 10^{-9}$& 0.9776 &  2.446   &   0.839   &   2.566& 0.8113965  \\
1  & $4.6461\cdot 10^{-9}$&  0.9861 &  3.060   &   0.941  &    2.254 & 0.8113434  \\
2  & $3.9363\cdot 10^{-9}$ & 0.9856 &  2.389   &   0.928   &   1.833 & 0.8113753  \\
5  & $\mathbf{3.4729\cdot 10^{-9}}$&  \textbf{0.9873} &   1.970  &   0.895   &    1.578& 0.8113620  \\
10  & $4.0484\cdot 10^{-9}$&  0.9856 &   2.569 &   1.002   &    1.972& 0.8113434  \\
50  & $1.3790\cdot 10^{-8}$&  0.9601 &   14.672  &   2.154   & 9.908 &  0.8113262   \\
100 &$2.4734\cdot 10^{-8}$& 0.9174  &   18.587  &   2.450   &  12.446 & 0.8112975  \\
\hline
\hline

\end{tabular}
\end{table}
\begin{table}[h!]
\centering
\caption{The MSE and NCC between the model and reconstructions of dataset C using different values of $\mu$. The optimal values of the MSE and NCC are indicated in bold. The data-likelihood term of the final image $f_{data}$, its components per type of observable (over the number of that observable type.) and the "regularization" term $-\log(D(x_f))$ are also listed.}
\label{ModelComparisonC}
\begin{tabular}{ c  c  c c c c c} 
\hline
\hline
$\mu$  & MSE & NCC & $\chi^2_{v^2}/N_{v^2}$ & $\chi^2_{c.p.}/N_{c.p.}$ & $f_{data}$& $-\log(D(x_f))$\\
\hline
$10^{-6}$  & $4.1767\cdot 10^{-9}$&   0.9746 &  1.016 &   1.006   &   1.012 & 0.8113182 \\
0.0001  & $4.2596\cdot 10^{-9}$&  0.9743 & 1.015    &   1.003   &    1.010&  0.8113189\\
0.001  & $3.9964\cdot 10^{-9}$&  0.9757 & 1.007   &   0.942  &   0.982&  0.8113192 \\
0.01  & $3.5799\cdot 10^{-9}$&  0.9785 & 0.994   &   0.933  &    0.971& 0.8113183 \\
0.1  & $3.6523\cdot 10^{-9}$& 0.9779 &  0.935   &   1.006  &   0.980& 0.8113186  \\
1  & $3.1727\cdot 10^{-9}$&  0.9826 &   0.990  &   0.891  &    0.952& 0.8113150 \\
2  & $\mathbf{2.5417\cdot 10^{-9}}$ & \textbf{0.9864} &  0.990   &   0.895   &   0.954& 0.8113124   \\
5  & $2.7034\cdot 10^{-9}$&  0.9847 &   1.017  &   0.907   &    0.976&  0.8113014 \\
10  & $4.0376\cdot 10^{-9}$&  0.9734 &   1.334 &   1.783   &    1.504& 0.8112868\\
50  & $4.0191\cdot 10^{-8}$&  0.6847 &   4.962  &   17.167   &  9.580 &  0.8112635  \\
100 &$5.5828\cdot 10^{-8}$& 0.5341  &   6.369  &   22.045   &  12.300 &  0.8112569 \\
\hline
\hline
\end{tabular}
\end{table}

%\begin{table}[h]
%\centering
%\caption{The MSE and  between the model and reconstructions of data with Gaussian using different values of $\mu$. }
%\label{ModelComparisonB}
%\begin{tabular}{ c  c  c c c c } 
%\hline
%\hline
%$\mu$  & MSE & NCC & $\chi^2_{v^2}/N_{v^2}$ & $\chi^2_{c.p.}/N_{c.p.}$ & $f_{data}$\\
%\hline
%$10^{-6}$  & $3.1609\cdot 10^{-9}$&   0.9802 & 0.989  &   1.055  &   1.014   \\
%0.0001  & $3.3374\cdot 10^{-9}$&  0.9788 & 0.997   &  1.164   &    1.060  \\
%0.001  & $3.5284\cdot 10^{-9}$&  0.9778 & 1.002   &   1.071   &    1.028   \\
%0.01  & $3.4378\cdot 10^{-9}$&  0.9785 & 0.989    &   1.130  &    1.043   \\
%0.1  & $3.1315\cdot 10^{-9}$& 0.9808 &  0.986   &   1.014   &   0.997  \\
%1  & $2.5497\cdot 10^{-9}$& 0.9857 &  0.979   &   0.963  &    0.973   \\
%2  & $2.6781\cdot 10^{-9}$ & 0.9852 &  0.991   &   0.954   &   0.976    \\
%5  & $3.1082\cdot 10^{-9}$&  0.9811 &   1.970  &   0.895   &    1.578   \\
%10  & $5.7124\cdot 10^{-9}$&  0.9613 &   1.424 &   3.823   &    2.332   \\
%50  & $4.7487\cdot 10^{-8}$&  0.6163 &   4.883  &   18.481   & 10.028      \\
%100 &$6.0789\cdot 10^{-8}$& 0.4750  &   6.385  &   23.936   &  13.026   \\
%\hline
%\hline
%\hline
%\end{tabular}
%\end{table}

\clearpage
\section{UV-coverages}\label{UVcov}
\begin{figure*}[th]
    \begin{subfigure}{0.5\linewidth}
    \includegraphics[width=\textwidth]{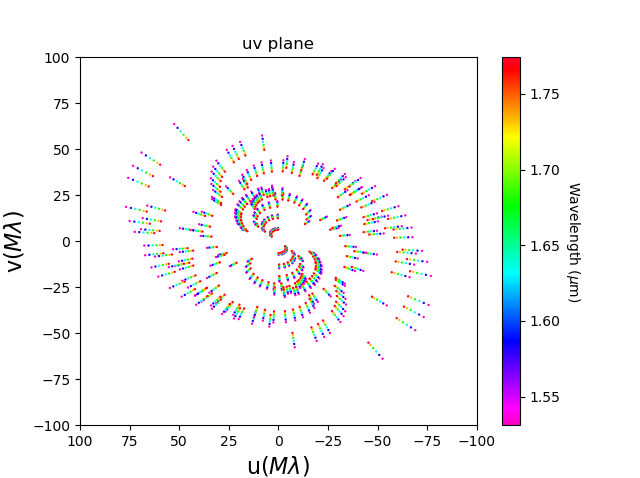}
    \end{subfigure}
    \qquad
    \begin{subfigure}{0.5\linewidth}
    \includegraphics[width=\textwidth]{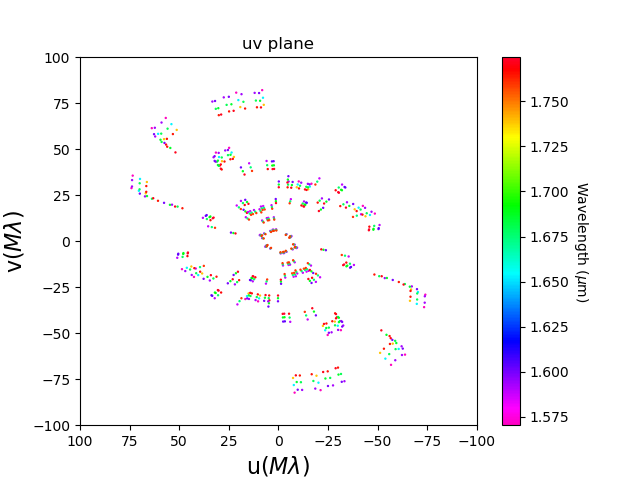}
    \end{subfigure}
    \caption{The UV-coverges of IRAS08544-4431 (left) and HD456777 (right).}
    \label{ArtificialObservablesA}
\end{figure*}

%\section{L-curves}\label{L-curves}
%\begin{figure*}[h!]
%    \begin{subfigure}{0.5\linewidth}
%    \includegraphics[width=\textwidth]{Artificial_Datasets/Lcurve_Artificial_IRAS.png}
%    \end{subfigure}
%    \qquad
%    \begin{subfigure}{0.5\linewidth}
%    \includegraphics[width=\textwidth]{Artificial_Datasets/Lcurve_HD_artificial.png}
%    \end{subfigure}
%    \caption{The L-curve for artificial dataset A (left) and B (right). The colors indicate the value of $\mu$.}
    %[9.0_-0.6_0.5]\ImageAI_i25.0_l00001.50
%    \label{ArtificialObservablesA}
%\bigskip
%    \includegraphics[width = 0.5\textwidth]{Artificial_Datasets/L-curveIWCar.png}
%    \caption{The L-curve for artificial dataset C. The colors indicate the value of $\mu$.}
%    \label{fig:my_label}
%\end{figure*}
\section{Comparison between data and reconstruction}\label{artificialData}
\begin{figure}[h]
    \centering
    \includegraphics[width=\textwidth]{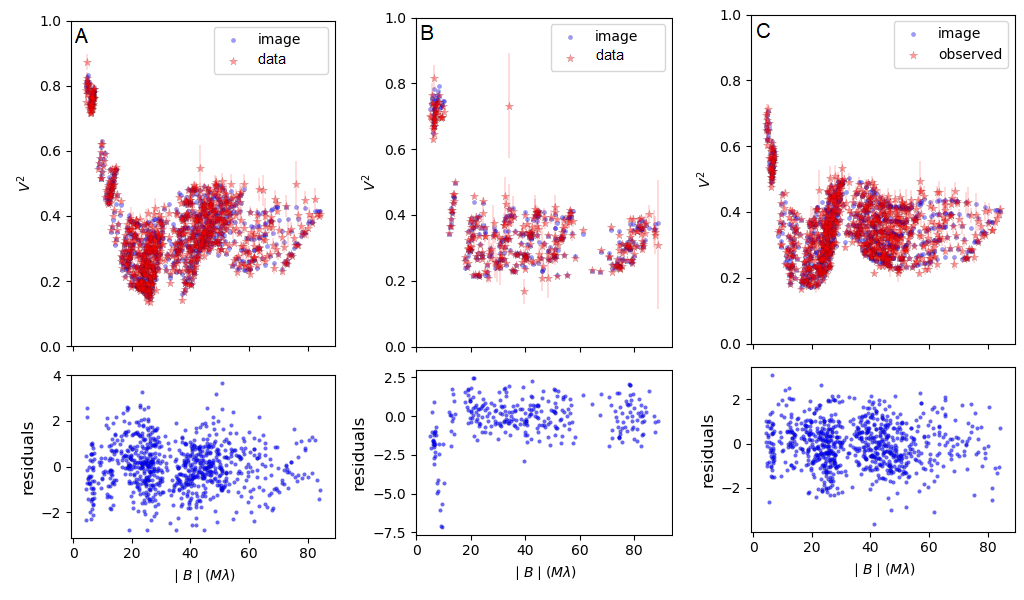}
    \caption{Top: Comparison between the squared visibilities of the artificial datasets and the corresponding best image reconstruction. The contributions of a central star are present in both. Bottom: the residuals normalized by the error on the corresponding data point. The residuals are given as $(v^2_{data}-v^2_{image})/\sigma_{v^2}$.}
    \label{fig:my_label}
\end{figure}
\begin{figure}[h]
    \centering
    \includegraphics[width=\textwidth]{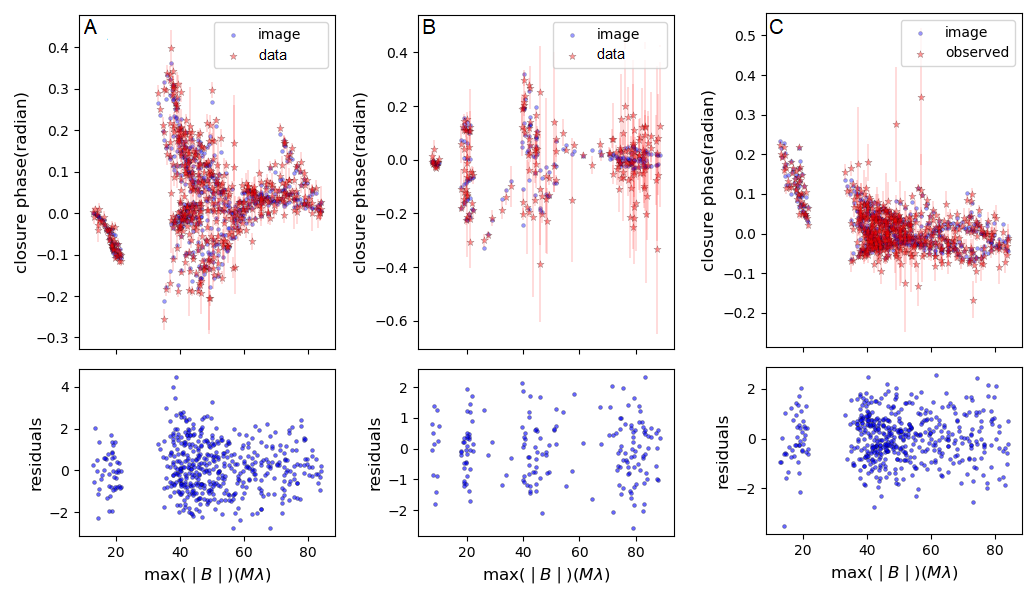}
    \caption{Top: Comparison between the closure phases of the artificial datasets and the corresponding best image reconstruction. The contributions of a central star are present in both. Bottom: the residuals normalized by the error on the corresponding data point. The residuals are given as $(cp_{data}-cp_{image})/\sigma_{cp}$.}
    \label{fig:my_label}
\end{figure}
%\begin{figure*}[t]
%    \begin{subfigure}{0.49\linewidth}
%    \includegraphics[width=\textwidth]{Artificial_Datasets/Artificial_IRAS_V2.png}
%    \end{subfigure}
%    \qquad
%    \begin{subfigure}{0.49\linewidth}
%    \includegraphics[width=\textwidth]{Artificial_Datasets/Artificial_IRAS_CP.png}
%    \end{subfigure}
%    \caption{The Squared visibilities (Right) and closure phases (Left) of artificial dataset A.}
%    \label{ArtificialObservablesA}
%\end{figure*}

%\begin{figure*}[t]
%    \begin{subfigure}{0.49\linewidth}
%    \includegraphics[width=\textwidth]{Artificial_Datasets/Artificial_HD_V2.png}
%    \end{subfigure}
%    \qquad
%    \begin{subfigure}{0.49\linewidth}
%    \includegraphics[width=\textwidth]{Artificial_Datasets/Artificial_HD_CP.png}
%    \end{subfigure}
%    \caption{The Squared visibilities (Right) and closure phases (Left) of artificial dataset B.}
%    \label{ArtificialObservablesA}
%\end{figure*}
%\begin{figure*}[t]
%    \begin{subfigure}{0.49\linewidth}
%    \includegraphics[width=\textwidth]{Artificial_Datasets/added_Gaussian_V2.png}
%    \end{subfigure}
%    \qquad
%    \begin{subfigure}{0.49\linewidth}
%    \includegraphics[width=\textwidth]{Artificial_Datasets/added_Gaussian_CP.png}
%    \end{subfigure}
%    \caption{The Squared visibilities (Right) and closure phases (Left) of artificial dataset Gaussian!!!!!!.}
%    \label{ArtificialObservablesA}
%\end{figure*}
%Three artificial datasets where created in order to validate our image reconstruction method. the closure phases of these artificial datasets can be seen 
\newpage
\acknowledgments % equivalent to \section*{ACKNOWLEDGMENTS}
We would like to thank Jeroen Audenaert, Ferréol Soulez and Giuseppe Marra for interesting discussions on the work leading to this paper.
RC, JK and HVW acknowledge support from the research council
of the KU Leuven under grant number C14/17/082.
The resources and services used in this work were provided by the VSC (Flemish Supercomputer Center), funded by the Research Foundation - Flanders (FWO) and the Flemish Government.
This research has made use of the Jean-Marie Mariotti Center OiDB service available at http://oidb.jmmc.fr.
% References

\bibliography{report} % bibliography data in report.bib
\bibliographystyle{spiebib} % makes bibtex use spiebib.bst
%!!!!!!!!!!!!!!!!! .bst file has been altered!!!!!!!
%following https://tex.stackexchange.com/questions/26575/bibtex-how-to-reduce-long-author-lists-to-firstauthor-et-al#:~:text=Put%20differently%2C%20this%20setup%20tells,has%20more%20than%20four%20authors.
\end{document}